# An innovative way of etching $MoS_2$: Characterization and mechanistic investigation

Yuan Huang[1,2,3,†], Jing Wu[2,3,†], Xiangfan Xu[2,3], Yuda Ho[2,3], Guangxin Ni[2,3], Qiang Zou[1], Gavin Kok Wai Koon[2,3], Weijie Zhao[2,3,4], A. H. Castro Neto[2,3,6], Goki Eda[2,3,4], Chengmin Shen[1] and Barbaros Özyilmaz[2,3,5,6]*

[1]Beijing National Laboratory for Condensed Matter Physics, Institute of Physics, Chinese Academy of Sciences, Beijing, China 100190
[2]Department of Physics, 2 Science Drive 3, National University of Singapore, Singapore 117542
[3]Graphene Research Centre, 6 Science Drive 2, National University of Singapore, Singapore 117542
[4]Department of Chemistry, National University of Singapore, 6 Science Drive 2, Singapore 117546
[5]Nanocore, 4 Engineering Drive 3, National University of Singapore, Singapore 117576
[6]NUS Graduate School for Integrative Sciences and Engineering (NGS), Centre for Life Sciences (CeLS), 28 Medical Drive, Singapore 117456.

## ABSTRACT

We report a systematic study of the etching of $MoS_2$ crystals by using $XeF_2$ as a gaseous reactant. By controlling the etching process, monolayer $MoS_2$ with uniform morphology can be obtained. The Raman and photoluminescence spectra of the resulting material were similar to those of exfoliated $MoS_2$. Utilizing this strategy, different patterns such as a Hall bar structure and a hexagonal array can be realized. Furthermore, the etching mechanism was studied by introducing graphene as an etching mask. We believe our technique opens an easy and controllable way of etching $MoS_2$, which can be used to fabricate complex nanostructures, such as nanoribbons, quantum dots and transistor structures. This etching process using $XeF_2$ can also be extended to other interesting two-dimensional crystals.

* Address correspondence to Barbaros Özyilmaz, barbaros@nus.edu.sg
†These two authors made an equal contribution to the work.

Two-dimensional (2D) crystals such as such as boron nitride (BN), bismuth telluride ($Bi_2Te_3$) and tungsten disulfide ($WS_2$) have attracted much attention due to their unique properties when exfoliated from the corresponding bulk 3D materials [1–4]. Graphene is by far the most studied 2D crystal because of its outstanding optical, mechanical, thermal and electronic properties [5–10]. However, pristine graphene does not have a band gap, which poses a major problem for its practical applications in high-performance field-effect transistors (FETs). Band gap engineering via nanoribbons or strain engineering brings about a deterioration of the unique transport properties in graphene [11, 12, 13]. Thus, searching for other 2D materials with suitable band gap is an important aspect from both scientific and application points of view.

$MoS_2$ is an indirect gap semiconductor with a 1.2 eV bandgap in its bulk form; single layer $MoS_2$, on the other hand, is a direct gap semiconductor with a 1.8 eV bandgap [14]. The indirect-to-direct transition arises from quantum confinement effects as thickness decreases [15], which also results in an enhancement of the photoluminescence of monolayer $MoS_2$ [16]. A single layer $MoS_2$ transistor with a room temperature current on/off ratio as high as $1 \times 10^8$ and ultralow standby energy dissipation has been reported [17], which makes it a promising candidate for use in semiconductor nanoelectronic devices. The excellent mechanical and optical properties of $MoS_2$ also make it a material [18, 19, 20], which can be used in applications where flexibility and transparency are required.

Currently, most $MoS_2$ devices have been made by using exfoliated flakes directly. This can be attributed to the chemical and physical properties of $MoS_2$, which make it impossible to react with common acids (such as hydrochloric acid, nitric acid and sulphuric acid) and bases (KOH, and NaOH) at room temperature [21, 22]. However, the ability to etch MoS$_2$ monolayer to give different geometries is critical for both fundamental physics studies, i.e., spin-Hall & spin-valley exploration [23] and also its potential applications. Thus, a new route to define sophisticated structures of MoS$_2$ is in high demand.

In this paper, we report a chemical dry-etching and patterning method for MoS$_2$ by using XeF$_2$ as a gaseous reactant. Atomic force microscopy (AFM) was used to study the morphology before and after etching, and the etching rate was also calculated by comparing the thickness change. Raman spectroscopy, as a powerful tool to analyze structural and doping information, was also employed to study the changes after etching. The photoluminescence (PL) spectra of single layer MoS$_2$ after etching and exfoliated single layer were compared. In order to gain a deeper insight into the etching mechanism, chemical vapor deposition (CVD) graphene was used as an etching mask. Combined with traditional micro–nanofabrication technology, where poly (methyl methacrylate) (PMMA) was used as etching mask after electron beam lithography patterning, we successfully obtained a hexagonal MoS$_2$ pits array after transferring exfoliated graphene as a mask. This is the first systematic report of such etching of MoS$_2$, which will be useful for future investigations and applications.

MoS$_2$ and graphite crystal were purchased from RS Components. Single layer CVD graphene grown on copper foil was supplied by Samsung Company. XeF$_2$ with purity of 99.9% was used as etching gas in our experiments. An optical microscope (Zeiss imager. A 1m, RIC Facility) was used to image MoS$_2$ flakes. The etching process was carried out in a xenon difluoride etching system (e1 Mod, XACTIX). Morphologies of MoS$_2$ samples were examined by AFM (Dimension FastScan, Bruker). The Raman spectra and mappings were collected by using a Raman Microscope (Alpha 300R,

WITec). Photoluminescence spectra were observed by confocal fluorescence microscopy (NETGRA Spectra, NT-MDT). Few-layer exfoliated graphene was transferred onto $MoS_2$ flakes using a home-made transfer system. An e-beam lithography system (Nova NanoSEM 50, FEI Corp.) and a reactive ion etching system (VITA-MINI) were used for patterning the array of graphene holes.

$MoS_2$ multilayer flakes were obtained by mechanical exfoliation of bulk $MoS_2$ on a silicon substrate covered with 285 nm of thermally oxidized $SiO_2$. The substrate surface was cleaned in the RIE system before exfoliating flakes on it. When the system was pumped down to $1.0 \times 10^{-3}$ Pa, oxygen gas was introduced at a flow rate of 20 sccm (sccm denotes cubic centimeters per minute at STP) to generate an oxygen plasma; the cleaning process takes 5 min at 20 W.

CVD graphene grown on copper foil was spin-coated with one layer of PMMA, after etching the copper foil with $FeCl_2$ solution, PMMA and graphene floated up. Thus we could easily transfer graphene onto the $SiO_2$ substrate with $MoS_2$ flakes. PMMA was removed by immersing in acetone for 3 hours.

In order to transfer exfoliated graphene onto $MoS_2$ flakes, we used the same mechanical transfer method as used in the fabrication of graphene/h-BN heterostructures [24]. We used an optical mask in this transfer process, which includes three layers: a glass slide, adhesive tape and one layer of PMMA. All of these three layers are transparent, so we could easily align the position of $MoS_2$ and graphene. After preparing the mask, we exfoliated graphene on top of PMMA, and then transferred few-layer graphene under an optical microscope.

Due to its highly oxidizing properties, $XeF_2$ is widely used in industry as an etching gas for Si. Recently, XeF$_2$ has also been used as a powerful tool to modify the properties of graphene, such as conductivity, transparency and bandgap [25, 26, 27]. We chose XeF$_2$ for two reasons: firstly, due to its highly oxidizing properties, the gas reacts easily with MoS$_2$; secondly, the reaction by-products can be efficiently removed as gases, such as Xe, SF$_6$, F$_2$, and MoF$_3$ et al. The reaction can be given as follow:

$$MoS_2 + XeF_2 \longrightarrow Xe + MoF_4 + SF_6$$
$$MoF_4 \longrightarrow MoF_3 + F_2$$

This strong oxidation-reduction reaction produces a large amount of heat and the temperature of the MoS$_2$ flakes is like to increase as a consequence, thus resulting in a positive influence on the reaction rate. All the products of this reaction are gases at room temperature. So at the end of the etching we can get very clean interface on our test samples.

The optical contrast on the substrate is sufficient to identify MoS$_2$ flakes with thickness down to a monolayer [17]. Figure 1a shows an optical image of a multilayered MoS$_2$ flake on Si/SiO$_2$ substrate. Figure 1b shows the same flake as in Fig. 1a after the etching process. The red zone marked in Fig. 1a was completely etched, whilst the remainder of the flakes remained due to their higher thickness.

The AFM images of one MoS$_2$ flake were also acquired on a 200 × 200 nm$^2$ scan window, and are shown in Figs. 1c and 1d. As clearly seen in these two images, the surface morphology of this flake changed considerably after etching, and the root-mean-square (rms) surface roughness changed from 0.07 nm to 0.9 nm, which is in consistent with a previous report by Castellanos-Gomez et al. [28]. There are two possible reasons for this result. Firstly, the reaction might not be uniform on the surface because of some molecular adsorption and traces of un-removed MoS$_2$. It may also originate from the microstructure of MoS$_2$, whereby due to the sandwich structure S–Mo–S, the reaction rates of S layers and Mo layers might be different, thus causing an etching fluctuation.

In this study, all the etching experiments were conducted at room temperature. Higher pressures or longer times produced thinner flakes. Etching rate is very important for precise control of the etching thickness. In order to measure the etching rate, we prepared 24 samples at different etching times, and selected two different flakes in each sample for testing. The thickness of each flake was measured by AFM before and after etching. Figures S-1a and S1-b in the Electronic Supplementary Material (ESM) show the AFM images and height profile of the same $MoS_2$ flake, from which it can be seen that the thickness decreased by 18.8 nm after etching. The variation of etched thickness versus time can be clearly seen in Fig. 2; there was no obvious thickness change in the first 20 s, and the etching rate then accelerated after 30 s.

Although mechanical exfoliation methods have proven to be effective in giving high quality single or few layers $MoS_2$, most of the samples are very small and the yield is hard to control. Therefore, future research and applications require the development of a new procedure to precisely control the quality and layer numbers. Laser-thinning of $MoS_2$ has been reported as an effective way to get single-layer 2D crystals from multilayered flakes [28], but this method is still hard to control. After we calculated the etching rate, we also tried to obtain single layer $MoS_2$ by $XeF_2$ etching. Figure 3a shows an optical image of a multilayered $MoS_2$ flake on the substrate. As seen clearly in the marked zone, the flake was a golden color. Figure 3b shows the optical image of the same flake after $XeF_2$ etching for 120 s at 1 torr; one part of the marked zone became blue and one part became more transparent. AFM measurements showed that the edge thickness was 0.9 nm (Fig. S-2 in the ESM), which is consistent with a previous report [28].

Raman spectroscopy is a sensitive characterization method that has been widely used to study 2D materials [29, 30, 31]. For graphene, the layer numbers, defects, strain and substrate effects can be easily determined from its Raman spectrum [29, 32]. Lee et al. reported a systematic study of the Raman spectrum of $MoS_2$; the Raman shift changed gradually as the layer number increased [33]. The intrinsic peaks of $MoS_2$, $E^1_{2g}$ and $A_{1g}$ occur at around 380 and 400 $cm^{-1}$ respectively. The frequency difference between the two most prominent peaks can be used as a direct proof of layer number; however the difference remains almost the same for layer number larger than 5. In Fig. 3c, we compare the difference in Raman spectra before and after etching. The two peaks of thick $MoS_2$ flakes are located at 384 and 409 $cm^{-1}$, with the separation between the two peaks being about 25 $cm^{-1}$. As the thickness decreased after $XeF_2$ treatment, the frequency difference between the two peaks decreased to 19 $cm^{-1}$. The frequency difference between $E^1_{2g}$ and $A_{1g}$ is the same as for exfoliated monolayer $MoS_2$. This means that the intrinsic vibration mode did not change significantly even after etching. Spatial maps (5 μm × 5 μm) of the Raman frequency of modes $E^1_{2g}$ and $A_{1g}$ are compared in Fig. S-3 (in the ESM).Due to the transition from an indirect to a direct-bandgap semiconductor, single layer $MoS_2$ exhibits a unique signature in its photoluminescence spectrum. In Fig. 3d, we compare the PL spectrum of the etched single layer and one exfoliated single layer sample. The exfoliated single layer sample has two PL peaks at 670 nm and 619 nm, consistent with previous reports [34]. These two peaks are due to the direct excitonic transitions at the Brillouin zone K point [35]. The PL spectrum of the etched single layer proves that the etched single layer $MoS_2$ also retains the large intrinsic direct bandgap. However the intensity of the etched sample is larger than that of the pristine sample and the dominant peak at 663 nm, shows a blue shift which indicates a larger bandgap. The mechanism of this effect is still unknown, but it may arise from the quantum confinement effects as the thickness decreases. In addition, the exfoliated monolayer retains the S–Mo–S sandwich structure, but the etching-thinned monolayer sample could even break up the three-layer unit and make it into a thinner structure.

The etching process is closely related to the interaction between the bottom layer and the SiO$_2$ substrate. The reaction is an exothermic process, and elevated temperature can accelerate the reaction rate, such that the etching rate becomes faster as time increases as shown in Fig. 2. Because of the weak interlayer coupling between the MoS$_2$ layers, the heat cannot be easily dissipated to the surroundings during reaction. The bottom layer, however, remains due to the substrate effect. Similar effects have also been reported by other groups [28].

The fabrication of devices with different structures is critical for studying the intrinsic properties of 2D materials, such as Hall bar [7], FET [17, 36] and quantum dot devices [37]. By combining traditional electron beam lithography (EBL) and our etching method, we successfully fabricated some devices and patterns (see Fig. S-4 in the ESM). In these experiments, exfoliated MoS$_2$ flakes were put onto a SiO$_2$ substrate, and then one layer of PMMA was spin-coated on the chip. After EBL exposure and development, we obtained different patterns covered by PMMA. The patterned sample was put into the chamber and etched for 4 minutes at 1 torr pressure of XeF$_2$. Although etching with these parameters could completely remove MoS$_2$ flakes with a thickness less than 300 nm, the edge of the etched MoS$_2$ flake was still very sharp (Fig. S-4c in the ESM). This can be explained by forming one layer of PMMA film at the edge of MoS$_2$, because the heat during the reaction could make PMMA around the exposed area melt and cover the edge, and in this case the reaction will stop at the edge. Many groups have reported the physical properties of XeF$_2$-fluorinated graphene. Even though the properties and functionalities can change during the fluorination process, the 2D structure can be retained after fluorination [25, 26, 27]. This means graphene can be used as an etching mask for many materials, such as Si, SiO$_2$ and MoS$_2$. In order to study the etching mechanism, the stack of monolayer CVD graphene and MoS$_2$ flakes is etched for 3 minutes at 1 torr, as shown in Fig. 4a. Interestingly, from the AFM image we find that there are some hexagonal pits on each flake. As shown in Fig. 4c, there is a hole in the center of the hexagonal pits (Fig. 4d). The reason for this is that the many defects in CVD graphene provide reaction centers and the reaction by-products were removed as gas, while the covered area could be gradually etched from the defect centers. This means that the defects in CVD graphene can be easily detected in this way.

Considering that the position and amount of defects in CVD graphene is hard to control, exfoliated few-layer graphene is chosen as a mask for further study. Exfoliated graphene is an almost perfect 2D crystal, so the defects can be easily designed by using EBL and oxygen plasma. Figure 4b shows the optical image of the hexagonal pits array; the etched hole of graphene can be clearly seen at the center of every hexagonal trench and these trenches are aligned to each other. The main difference between PMMA and graphene as an etching mask is that graphene with its high thermal conductivity and stability does not form an insulating layer, which is why we could obtain hexagonal pits under graphene.

Considering that XeF$_2$ molecules are the only reactive species in this process, and only the presence of point defects could induce hexagonal pits, we build a simple model to explain the etching mechanism. In the microstructure of monolayer MoS$_2$, each Mo atom is bonded to six S atoms, and each S atom is bonded to three Mo atoms (Fig. 5a).

Figure 5b shows the top view of a monolayer $MoS_2$ atomic structure, assuming that this monolayer $MoS_2$ was cover by CVD graphene with only one point defect in the center of the graphene. After we remove one S atom from the top layer (Fig. 5c), three Mo atoms in the middle layer and one S atom will be exposed to $XeF_2$, and they will become more active than the other atoms in the whole crystal. Then $XeF_2$ will selectively remove the four atoms, and in this case the first hexagonal hole can be observed (Fig. 5d). Following this, the exposed atoms in $MoS_2$ crystal will be removed in sequence, as shown in Figs. 5e–-5f, resulting in hexagonal pits after etching.

First-principles computation predicts that zigzag $MoS_2$ nanoribbons show ferromagnetic and metallic behavior, irrespective of the ribbon width and thickness, while the armchair nanoribbons are nonmagnetic and semiconducting [38，39]. Until now, there has been no experimental report of the properties of zigzag or armchair $MoS_2$ nanoribbons. As shown in our model, all the edges in the hexagonal pits are zigzag. This means zigzag $MoS_2$ nanoribbons can be fabricated by controlling the position of graphene defects and etching parameters, and its ferromagnetic properties could be realized on multilayered $MoS_2$ flakes.

**Conclusions**

We have discovered a strategy for etching $MoS_2$ using $XeF_2$ as a reaction gas. The morphology change and etching parameters were systematically studied. By controlling the etching process, monolayer $MoS_2$ with high quality can be realized. Raman and photoluminescence spectra demonstrated that the etched flakes are as regular as the pristine ones. By using PMMA as an etching mask, we successfully combined micro-fabrication technology with this etching method to obtain different geometric structures. In order to study the etching mechanism, graphene was employed as an etching mask, and the etching process was clearly illustrated by a simple model. The model indicates that zigzag MoS$_2$ nanoribbons can be fabricated by optimizing the etching process. Assuming that a wafer-scale single-crystal multilayered MoS$_2$ is available, this etching method could provide a useful tool for the fabrication of large-area single layer MoS$_2$ for integrated devices. Thus, a road can be paved for future fundamental studies and integrated nanodevices based on the method reported here.

## Acknowledgements

This work was supported by the NRF-CRP award "Novel 2D materials with tailored properties: beyond graphene" (R-144-000-295-281), the Singapore National Research Foundation Fellowship award NRF-RF2008-07, Singapore Millennium Foundation (SMF)-NUS Research Horizons Award 2009 (R-144-001-271-592, R-144-001-271-646), and the National Basic Research Program of China (973 Program, Grants No. 2013CB933604, No. 2011CB932703).

# Figure Caption

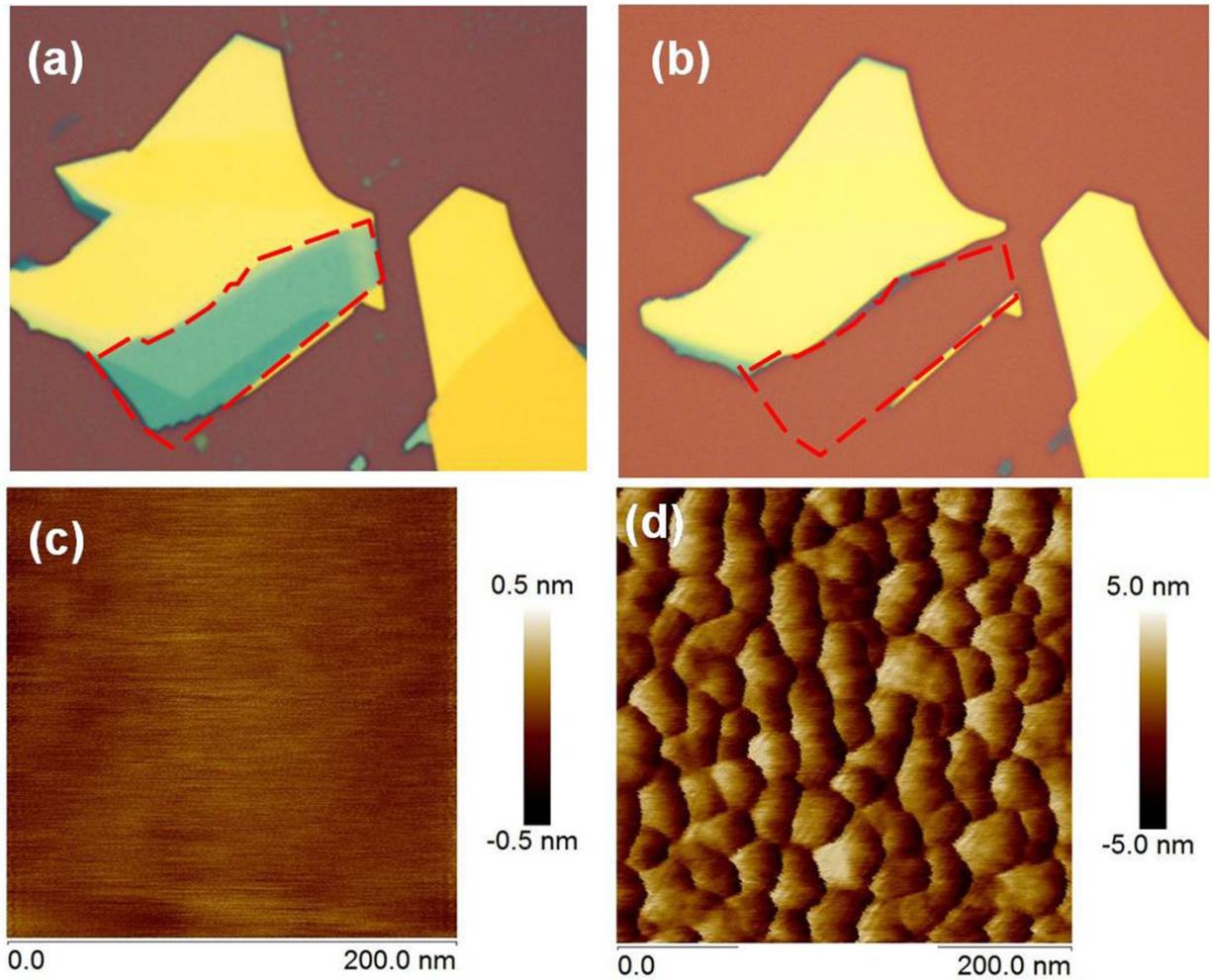

**Figure 1** Optical and AFM images of a $MoS_2$ flake. (a) and (b) are the optical images before and after etching. (c) and (d) are the corresponding AFM images of the lower right corner flake in (a) and (b).

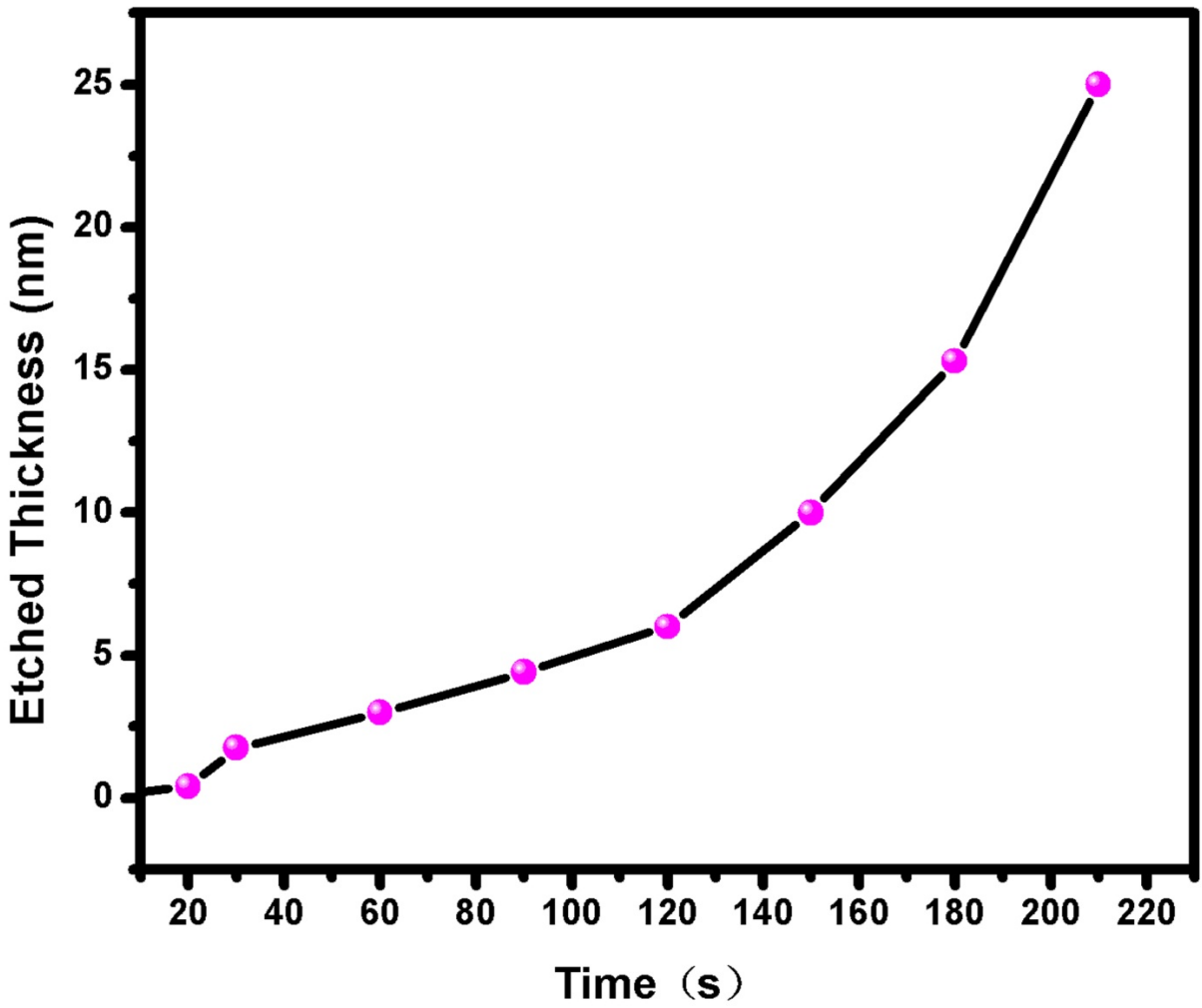

**Figure 2** Thickness measurement and etching rate analysis. Plot of etched thickness vs. time for 12 samples (the pressure was 1 torr, and each point corresponds to two samples).

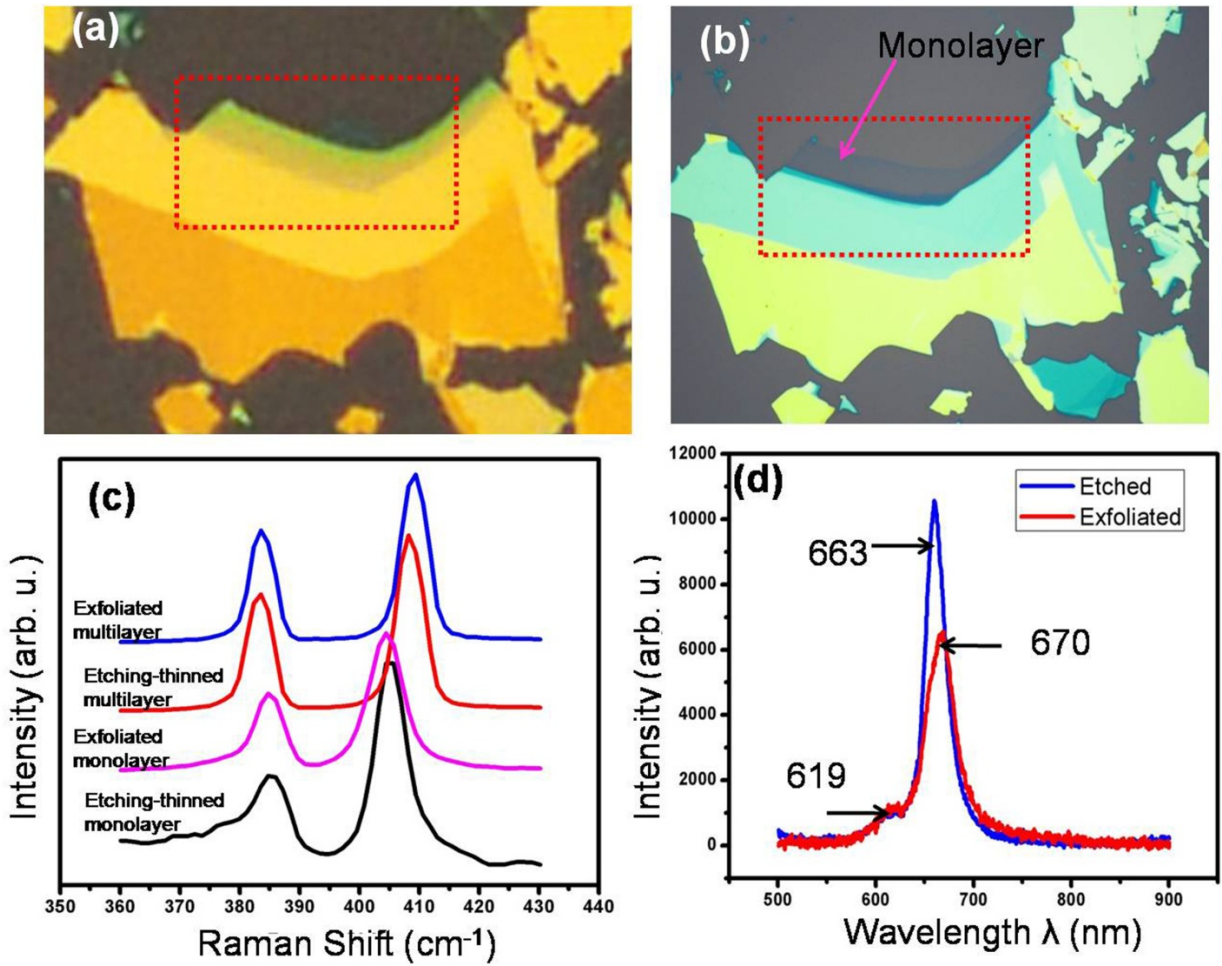

**Figure 3** Optical images of an etching-thinned monolayer and its photoluminescence spectrum. (a) Optical image of $MoS_2$ before etching. (b) Optical image of the same flake after etching; monolayer $MoS_2$ can be clearly seen in the marked zone. (c) Raman spectra of the exfoliated multilayer and monolayer and the corresponding etching-thinned samples. (d) The photoluminescence spectra of etching-thinned monolayer and exfoliated monolayer samples.

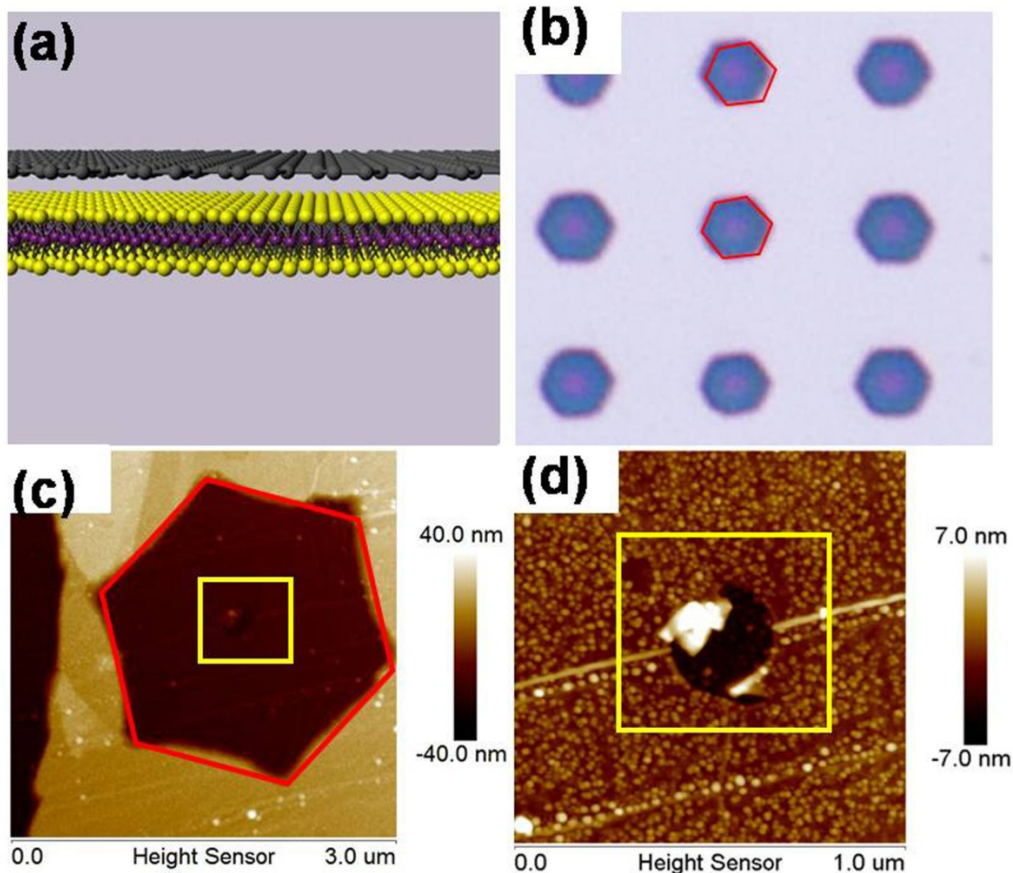

**Figure 4** AFM and optical images of hexagonal pits. (a) Schematic image of a MoS$_2$ flake after covering one layer of graphene. (b) Optical image of hexagonal array: few layer graphene was put onto this MoS$_2$ flake, then a hole array was patterned by combining EBL and oxygen plasma before etching. Two of the hexagonal pits are marked with red curves. (c) AFM image of the hexagonal pit, where the edge and center of the hexagonal pit are marked with red and yellow curves, respectively. (d) The zoom-in AFM image of the center in (c), which originated from a point defect of CVD graphene.

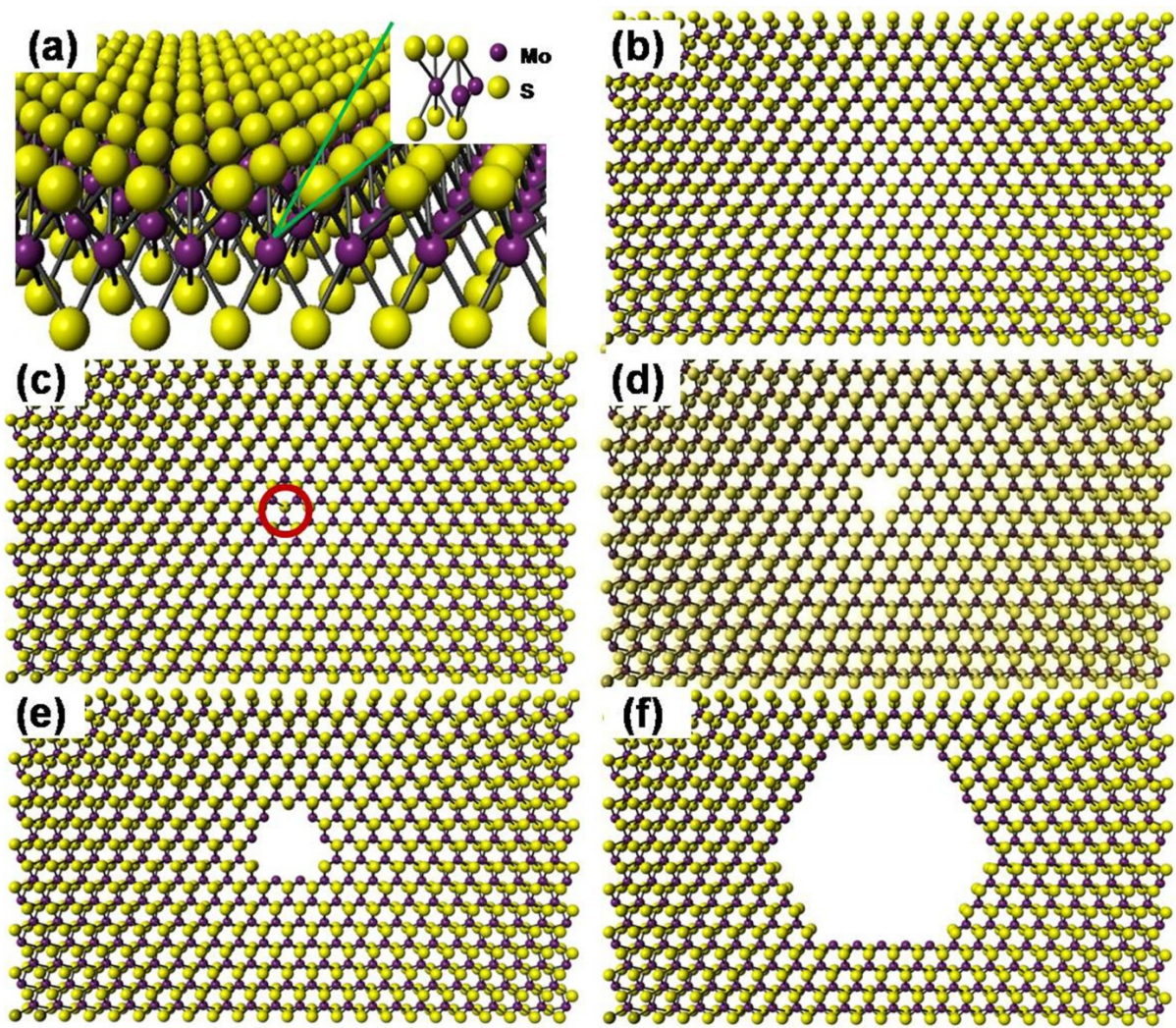

**Figure 5** Schematic model of the atomic structure of monolayer MoS$_2$ and the etching mechanism. (a) Atomic structure of monolayer MoS$_2$, where the inset shows the unit cell of the crystal. (b) Top view of a monolayer MoS$_2$ crystal, assuming the crystal was covered by CVD graphene. (c) After removing one S atom on top from one point defect of graphene (see the marked zone), three Mo atoms with dangling bonds are exposed to XeF$_2$. (d) After removing three Mo atoms in the middle layer and one S atom at the bottom by XeF$_2$. (e) and (f) The hexagonal pit growing larger during the etching process.

# Supplementary Material

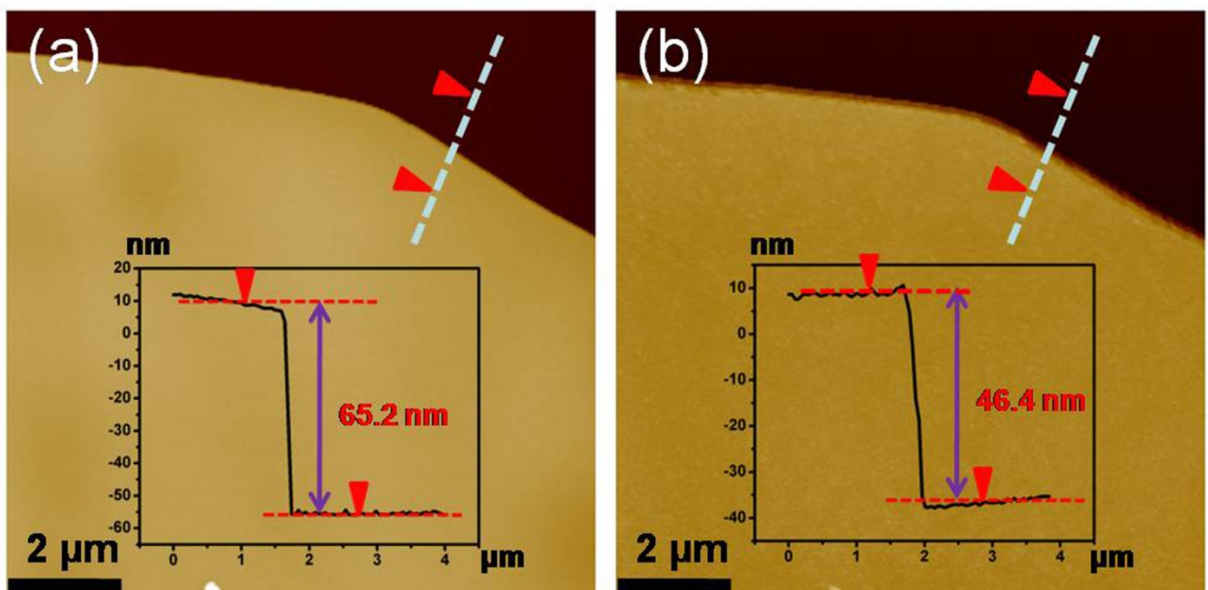

**Figure S-1** AFM images of the same MoS$_2$ flake before and after etching, with height profiles shown in the insets. (a) The thickness of pristine exfoliated MoS$_2$ is 65.2 nm. (b) The thickness of this flake decreased to 46 nm after etching for 180 s at 1 torr. By comparing the thickness changes, the plot of $\Delta T_{thickness}$ vs. time can be acquired as shown in Fig. 2.

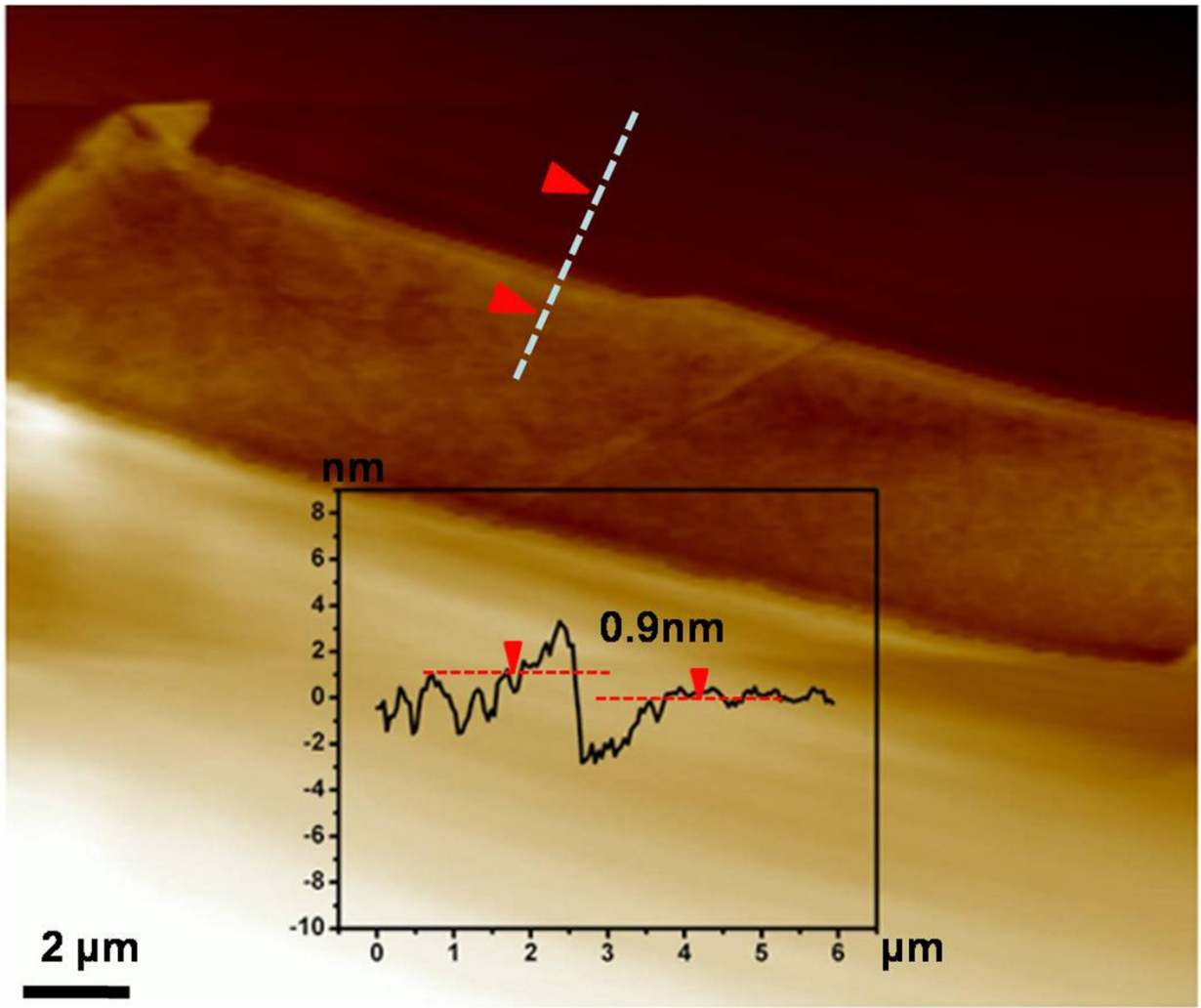

**Figure S-2** AFM image of the etching-thinned monolayer $MoS_2$, which is the sample shown in Fig. 3. The height profile is shown in the inset.

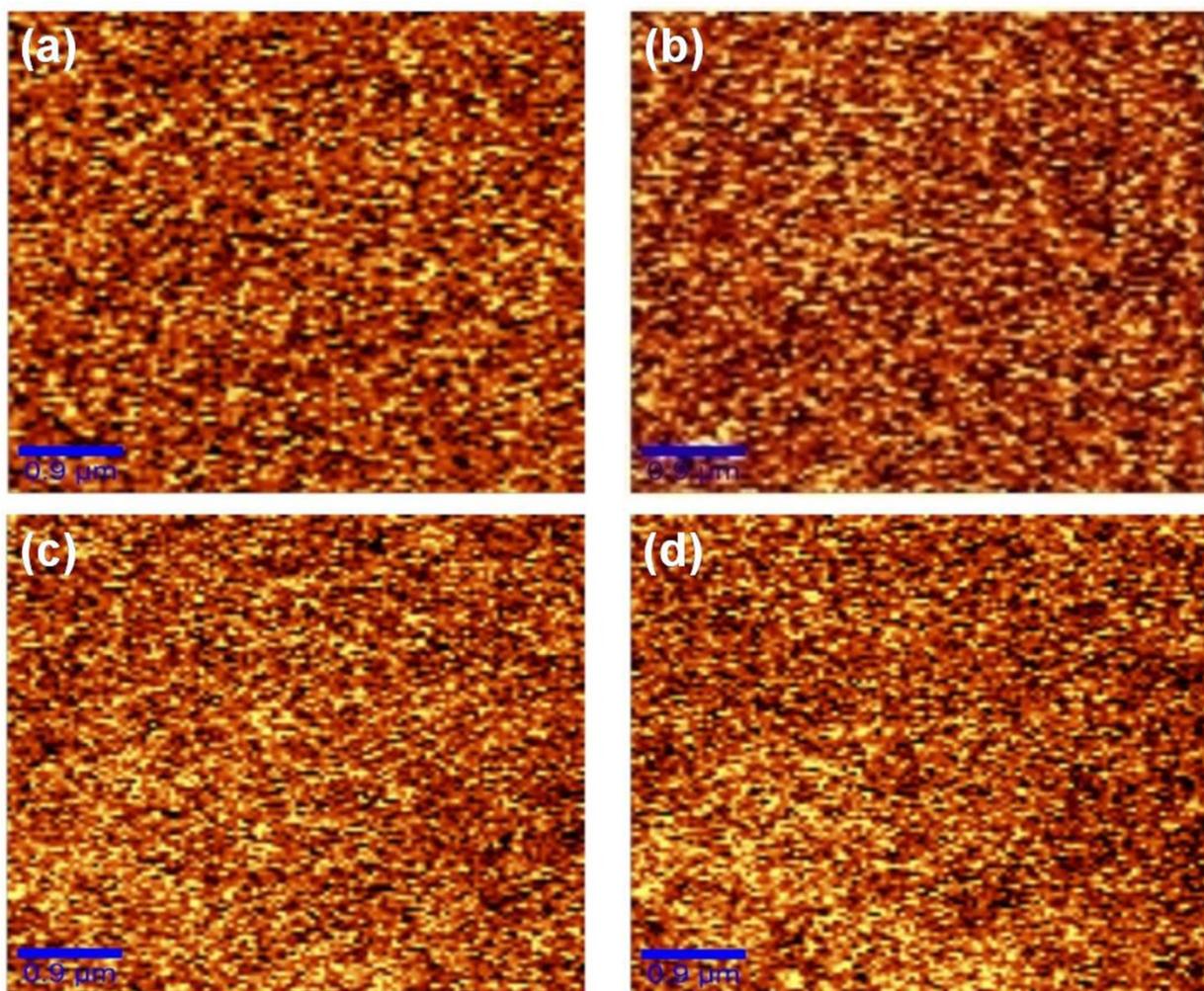

**Figure S-3** Spatial maps (5 μm × 5 μm) of Raman bands $E^1_{2g}$ (a, c) and $A_{1g}$ (b, d) for multilayered $MoS_2$. (a) and (b) are the spatial maps of an exfoliated $MoS_2$ flake, while (c) and (d) are the spatial maps of the same $MoS_2$ flake after etching.

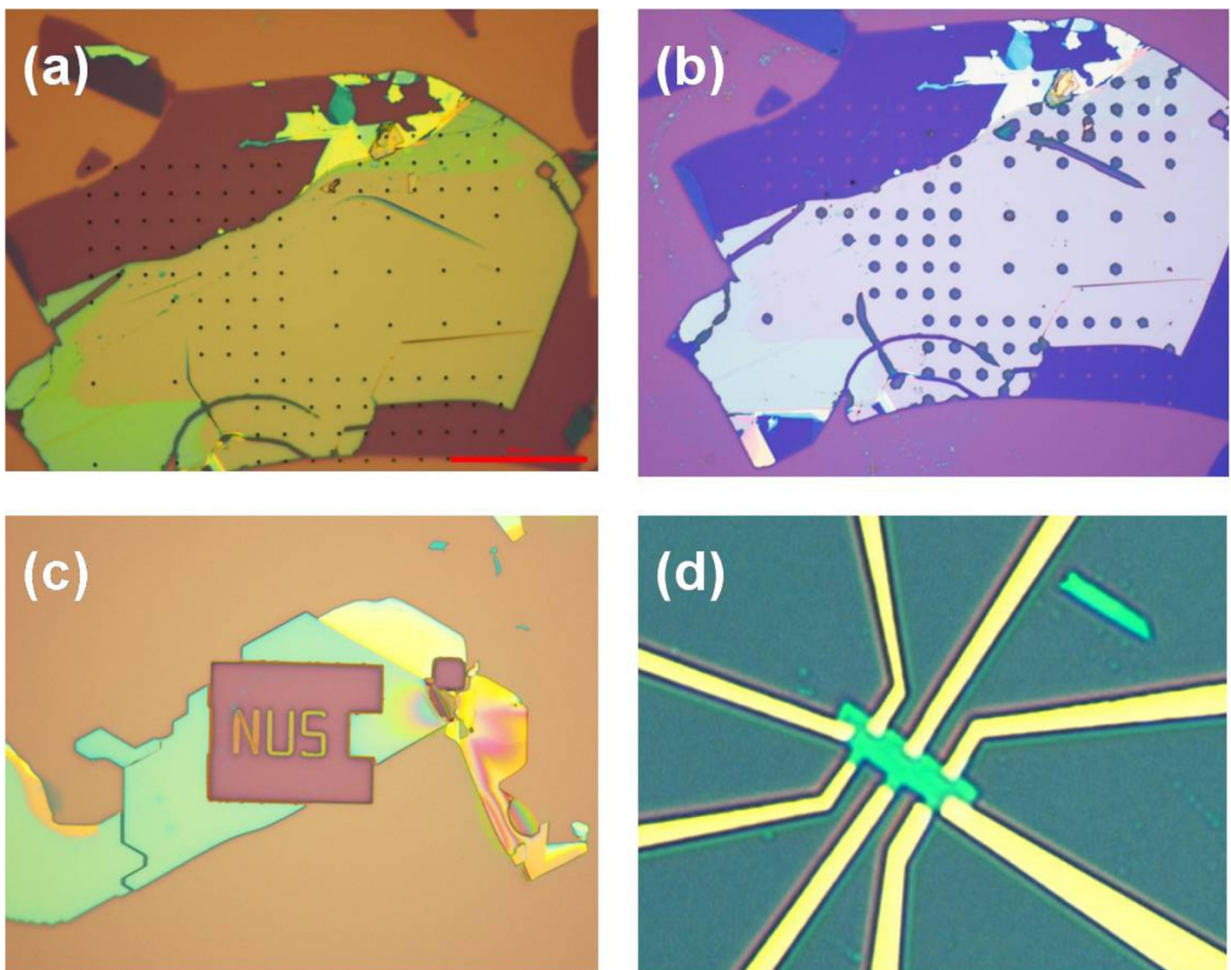

**Figure S-4** Images of MoS$_2$ pattern and device. (a) Image of MoS$_2$ flake covered by few-layer exfoliated graphene. An array of holes were fabricated on few-layer graphene by e-beam lithography and oxygen-plasma etching. Hexagonal pits grown in an MoS$_2$ flake from graphene holes after XeF$_2$ etching, as shown in (b) and Fig. 4b. (c) Image of one pattern on an MoS$_2$ flake. (d) Image of an MoS$_2$ device fabricated after e-beam lithography and XeF$_2$ etching.